\documentclass[lettersize,journal]{IEEEtran}
\usepackage{amsmath,amssymb,amsfonts}
\usepackage{algorithmic}
\usepackage{algorithm}
\usepackage{array}
\usepackage[caption=false,font=normalsize,labelfont=sf,textfont=sf]{subfig}
\usepackage{textcomp}
\usepackage{stfloats}
\usepackage{url}
\usepackage{verbatim}
\usepackage{graphicx}
\usepackage{cite}
\usepackage{color, xcolor}
\usepackage{lineno}

\usepackage{booktabs}
\usepackage{makecell}
\usepackage{multirow}

\makeatletter
\let\NAT@parse\undefined
\makeatother
\usepackage[colorlinks, linkcolor=black, citecolor=black, anchorcolor=black,urlcolor=black]{hyperref}  

\begin{document}

\title{Prior-Driven Self-Supervised Lightweight Method for Seismic Noise Attenuation}

\author{Junheng Peng, Yong Li, \IEEEmembership{Member, IEEE}, Yingtian Liu, Mingwei Wang

\thanks{Junheng Peng, Yong Li, Yingtian Liu, and Mingwei Wang are with the Key Laboratory of Earth Exploration \& Information Techniques of Ministry Education, and the School of Geophysics, Chengdu University of Technology, Chengdu, Sichuan, China.}
\thanks{Corresponding author: Yong Li(e-mail: Liyong07@cdut.edu.cn)}}



\maketitle

\begin{abstract}

Seismic exploration is currently the most mature approach for studying subsurface structures, yet the presence of noise greatly restricts its imaging accuracy. Previous methods still face significant challenges: traditional computational methods are often computationally complex and their effectiveness is hard to guarantee; deep learning methods rely heavily on datasets, and the complexity of network training makes them difficult to apply in practical field scenarios. In this paper, we proposed a neural network that has only 2464 learnable parameters, which is hundreds or even thousands of times lower than that of the current mainstream deep learning networks. And its parameter constraints rely on priors rather than requiring training data. We proposed two types of priors: the local prior and the global variance prior for self-supervised learning, and put forward low-scale learning to further enhance its performance in noise processing. We validated our method on both synthetic and field data, and the results indicate that our proposed approach effectively attenuates random noise. 

\end{abstract}

\begin{IEEEkeywords}
Seismic Noise Attenuation, Seismic Signal Processing, Image Processing
\end{IEEEkeywords}

\section{Introduction}

\IEEEPARstart{S}eismic exploration is currently one of the best methods for understanding the distribution of subsurface, playing a dominant role in industries such as oil and gas exploration. However, seismic exploration is highly susceptible to various forms of noise interference, resulting in decreased resolution.

Generally, conventional computational methods generally fall into three categories: filtering methods, sparse transform, and rank reduction. The precondition for using filtering methods is that seismic signals exhibit coherence or regularity, which can be distinguished from random noise in the frequency (F-X) domain \cite{8} and many authors have proposed improvement methods Building on this \cite{9}. Similar to filtering methods, sparse transform assumes that clean data exhibits structural characteristics that can be sparsely represented, while noise cannot due to its irregular and random nature \cite{11}. Common sparse transform include the Fourier \cite{12}, seislet \cite{13}, wavelet \cite{14}, and curvelet \cite{15}. In addition to using fixed orthogonal basis functions, another method capable of adaptively learning basis functions is referred as dictionary learning \cite{27}. The final method is rank reduction, which assumes that the seismic data is a low-rank structure, and the addition of random noise increases the rank of the data. Based on this, random noise can be eliminated by employing low-rank constraints \cite{16}. However, conventional computational methods have significant limitations: 1. Various filtering methods struggle to accurately separate noise from seismic data and often result in strong signal leakage; 2. Sparse transform and rank reduction suffer from low computational efficiency, especially when applied to large-scale exploration tasks; 3. Various conventional computational methods are relying on arbitrarily selected parameters, which introduces subjectivity into the processing. Recently, many authors have proposed using various deep learning methods for noise attenuation \cite{30}, such as: Zhao et al. \cite{18} proposed denoising CNN to learn noise extraction in blind Gaussian data; Liao et al. \cite{19} introduced the concept of iterative processing for ResNet \cite{20}, further enhancing the noise attenuation effectiveness; Peng et al. introduced diffusion model \cite{22} and fast diffusion model \cite{23} for processing seismic data with strong noise. However, various supervised deep learning methods face challenges in practical seismic exploration tasks, including: 1. Their effectiveness and generalization relies heavily on the training dataset; 2. They often lack interpretability. Another promising approach is to utilize unsupervised deep learning methods, such as: Liao et al. \cite{6} proposed twice DAE (TDAE) for seismic noise attenuation to mitigate both signals leaking and noise remaining exist; Liu et al. \cite{21} introduced weighted total variation (WTV) into the deep image priors, leveraging sparsity to promote priors and constrain the deep image priors solutions. The issue with the above methods lies in their complexity and low computational efficiency, making them impractical for large-scale seismic exploration tasks.

Therefore, in this study, we proposed a novel neural network for seismic noise attenuation, namely adaptive convolution filtering (ACF), which has a total of 2,464 learnable parameters and represents a reduction of several tens to hundreds of times compared to unsupervised deep learning methods (for example, TDAE has two DAE, and each DAE has learn more than 290,000 learnable parameters, which is more than 100 times that of ACF). ACF has a self-supervised deep learning technique and is entirely driven by data priors. We proposed using two priors to constrain ACF: local prior, global variance prior. The local prior effectively removes high-frequency random noise, and the global variance prior mitigates signal leakage by constraining the intensity of removed noise. In addition, in order to further improve the noise attenuation performance, we propose a low-scale learning strategy to deal with the residual noise. We conducted experiments on 2D and 3D seismic data, and the results demonstrate that ACF noise reduction outperforms several previous methods. Additionally, signal leakage remains at a reasonable level. Furthermore, to ensure the reproducibility of the experiments, we have made the code open-source.

\section{Methodology}

\subsection{Structure of Adaptive Convolutional Filter}

Inspired by some time-domain filtering methods, we propose using an extremely simple convolutional neural network for noise attenuation, which can be represented as follows:
\begin{equation}
M1^{h} = I \circledast C1_{k \times k \times 1}^h
\end{equation}
\begin{equation}
M2^{h} = M1^{h} \circledast C2_{k \times k \times h}^h
\end{equation}
\begin{equation}
O = M2^{h} \circledast C3_{1 \times 1 \times h}^1
\end{equation}
where $O$ represents the output seismic data, $I$ represents the noisy seismic data, $M1$ and $M2$ represents the feature maps, $C$ represents the convolutional kernel, and $\circledast$ represents convolution calculation. This is a simple three-layer convolutional structure, as shown in Figure~\ref{workflow}b, but it does not form a deep structure like deep learning networks. It is more like a filter rather than a deep learning network, so we call it an adaptive (parameter-learnable) convolutional filter. In this work, we used filter parameters where $k=3$ and $h=16$, with a total of 2464 learnable parameters. Compared to previous methods, its lightweight structure results in very low computational complexity. 

\subsection{Local Prior}

In our work, we employed the first prior proposed by Mansour et al. \cite{2} in their research. Firstly, the noisy seismic data $I$ is divided into its two subsamples through $I_1 = I \circledast k_1$, and $I_2 = I \circledast k_2$, where $k_1 = \begin{bmatrix} 0.5 & 0 \\ 0 & 0.5 \end{bmatrix}$ and $k_2 = \begin{bmatrix} 0 & 0.5 \\ 0.5 & 0 \end{bmatrix}$. The seismic data exhibit high correlation between each sampling point, whereas noise is unstructured and independent \cite{34}. Therefore, the two subsampled data can be regarded as two noisy observations of the same target \cite{35}, and based on this, the following constraint is defined: 
\begin{equation}
    \mathcal L_1(I) = \| ACF(I_1|\theta)-I_2 \|_2^2 + \| ACF(I_2|\theta)-I_1 \|_2^2
\end{equation}

However, the above constraints alone are insufficient, which may cause the ACF to overfit with some special values (such as all zeros). Therefore, to solve this, we have employed a symmetry constraint \cite{2}, which is expressed as follows:
\begin{equation}
\begin{aligned}
    \mathcal L_2(I) = \| ACF(I_1|\theta)-ACF(I|\theta)\circledast k_1 \|_2^2 \\ + \| ACF(I_2|\theta)-ACF(I|\theta)\circledast k_2 \|_2^2
\end{aligned}
\end{equation}
which can be regarded as a regularization term and can effectively avoid overfitting. In summary, the local prior can be expressed as:
\begin{equation}
    \mathcal L_{LP}(I)=\frac{1}{2} \times (\mathcal L_1+ \mathcal L_2)
\end{equation}

Although the local prior is very effective at eliminating noise, our experiments found that using this prior alone can cause significant signal leakage. Meanwhile, ACF is difficult to effectively remove residual noise (which is usually lower in frequency) based on $L_{LP}$ only. 

\subsection{Global Variance Prior}

In addition to the local prior, we proposed global prior for parameter constraints which allows global constraints to be achieved by constraining variance (which can also characterize the intensity of noise) of the noise removed by ACF, effectively improving the signal leakage caused by the locality of the local prior. 

In this study, we directly estimate the global variance $\sigma^2$ of noise $z$ from noisy seismic data $I$. First, since noise may not be uniformly distributed, we divide $I$ into multiple overlapping patches and unfold them into one-dimensional vectors $P = \{p_t\}_{t=1}^n$. If the size of the seismic data is $h\times w$ and the size of the patch is $d$, then $n$ is $(h-d+1)\times (w-d+1)$ and the dimension $r$ of $p_t$ is $d^2$. For $p_t$, it can be regarded as a combination of clean data $\hat{p}_t$ and random noise $z_t$. As assumed by many rank reduction methods, clean patches can be represented in a low-dimensional subspace $\hat{p}_t = Aq_t$, where $A \in \mathbb{R}^{r\times m}$ denotes subspace dictionary matrix with constraint $AA^\mathrm{T}=I$, $q_t\in \mathbb{R}^{m\times 1}$ denotes the projection of $\hat{p}_t$ in subspace and $m\ll r$. $A$ is typically solved using principal component analysis (PCA) and consists of the eigenvectors corresponding to the $m$ largest eigenvalues of the covariance matrix of $P$. 

Second, an additional matrix $U\in \mathbb{R}^{r\times (r-m)}$ is assumed, which together with matrix $A$ forms a new dictionary matrix $R = [A, U]$ with constraint $RR^\mathrm{T}=1$. Thus, $p_t$ can be rewritten as $p_t = R\begin{bmatrix}q_t, \\ o\end{bmatrix}+ z_t$. It should be noted that $R$ can be obtained by performing eigenvalue decomposition (such as PCA) on the covariance matrix of $P$. Multiplying $p_t$ by $R^\mathrm{T}$ yields $R^\mathrm{T}p_t =\begin{bmatrix}q_t + A^\mathrm{T}z_t, \\ U^\mathrm{T}z_t\end{bmatrix}$, where the target noise $z_t$ can be separated from the $(m+1)_{th}$ to the $r_{th}$ rows of the matrix $R^\mathrm{T}p_t$. We can further calculate the covariance matrix of $R^\mathrm{T}P$, which can ultimately be expressed as $\frac{1}{n}R^\mathrm{T}P(R^\mathrm{T}P)^\mathrm{T} = R^\mathrm{T}\frac{1}{n}PP^\mathrm{T}R = R^\mathrm{T}\Sigma R = \lambda$, where $\Sigma$ and $\lambda$ denote the covariance matrix of $P$ and its eigenvalue matrix, respectively. The matrix $\lambda$ is a diagonal matrix, and the $t_{th}$ eigenvalue corresponds to the variance of the $t_{th}$ row in $R^\mathrm{T}p_t$. According to the lemmas proposed by Chen et al. \cite{1}, we successively removed the largest eigenvalue from the sorted sequence of eigenvalues until the remaining eigenvalues satisfy the condition that the median equals their mean, and the expectation of the residual eigenvalue sequence represents the variance $\sigma^2$ of the noise $z$.

The $\sigma^2$ estimated in this way reflects the overall statistical feature of the noise rather than local features. In this work, we mainly use the noise variance constrain as follows:
\begin{equation}
    \mathcal L_{GVP}(I) = | \frac{1}{n}\sum_{i=1}^{i=n}(z'_i-Mean(z'))^2-\sigma^2|
\end{equation}
where $O$ belongs to $z'$, with $z'$ being the noise predicted by the ACF as $I - O$. 

\subsection{Low-Scale Learning}

Considering the local prior and global variance prior, we proposed the following loss function to continuously optimize the ACF parameters:
\begin{equation}
    \mathcal L(I) = \frac{1}{2} \times (\mathcal L_{LP}(I)+ \mathcal L_{GVP}(I))
\end{equation}

\begin{figure}
\centering
\noindent\includegraphics[width=8cm]{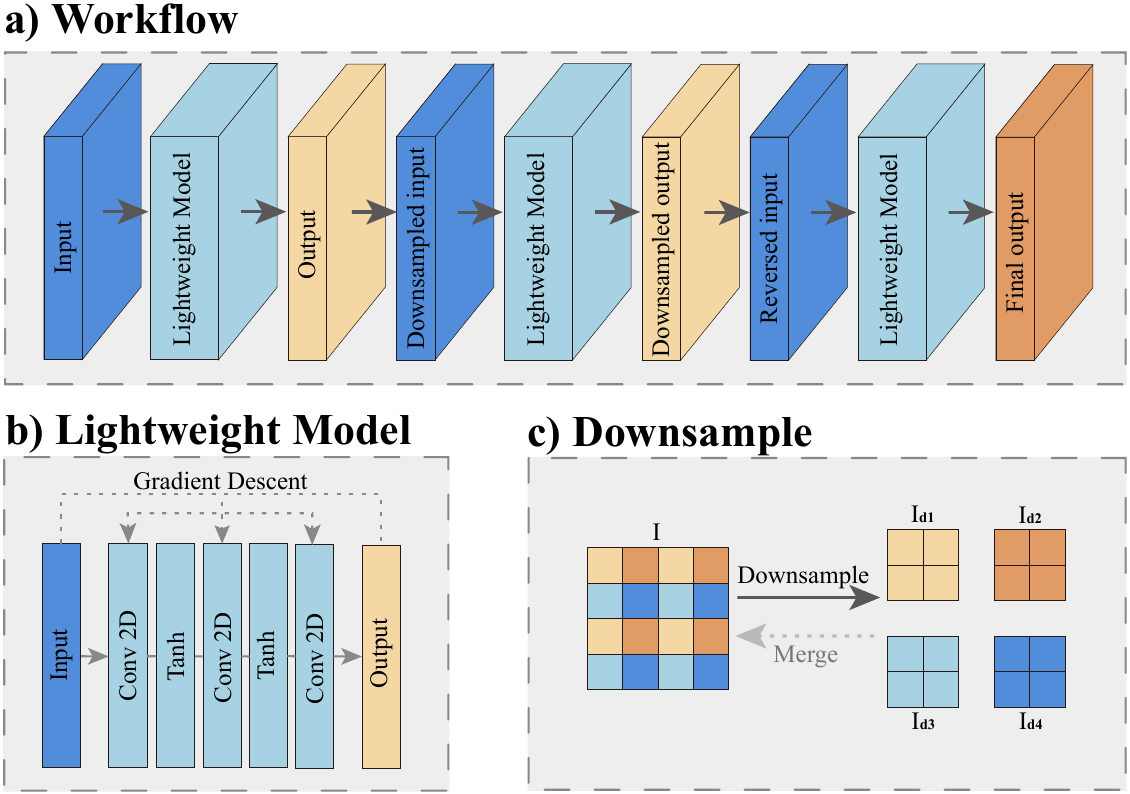}
\caption{a) The workflow of our proposed method; b) the structure of ACF; c) the downsampling and merging methods of downsampling learning.}
\label{workflow}
\end{figure}

Typically, seismic data exhibit large-scale structures, whereas noise either lacks or only has low-scale structures, while downsampling the data will destroy low-scale structures. Therefore, to further attenuate noise, we propose a low-scale learning method. We interval-sample the seismic data containing residual noise to generate downsampled data $\{I_{dn}\}_{n=1}^4 = I \circledast d_n$, where $d_1 = \begin{bmatrix} 1 & 0 \\ 0 & 0 \end{bmatrix}$, $d_2 = \begin{bmatrix} 0 & 1 \\ 0 & 0 \end{bmatrix}$, $d_3 = \begin{bmatrix} 0 & 0 \\ 1 & 0 \end{bmatrix}$, and $d_4 = \begin{bmatrix} 0 & 0 \\ 0 & 1 \end{bmatrix}$, as shown in Figure~\ref{workflow}c. Since multiple low-scale data after downsampling exhibit high similarity, we use the same ACF for their processing, and its optimization satisfies: 
\begin{equation}
ACF(\theta)=\mathrm{arg}\min_{\theta} \sum_{i=1}^{4} \mathcal L(I_{di})
\end{equation}

After obtaining the processed results, the downsampled data are merged. The merged data inevitably contains a small amount of noise due to boundary effects caused by slight signal leakage. Therefore, we utilized ACF again to remove this minimal noise, resulting in the final output. Based on the aforementioned priors and constraints, the overall process of the proposed ACF is illustrated in Figure~\ref{workflow}a. Based on the previously proposed loss function, the parameters are continuously updated using gradient descent to achieve adaptive seismic data noise attenuation without the need for additional data or setting of parameters. Although our proposed workflow requires three updates of the ACF parameters, its lightweight structure and training strategy can still effectively ensure overall computational efficiency.

\section{Examples}
\subsection{Synthetic Data Examples}

In the experimental section, since perfectly clean field data is typically unavailable, we first evaluated the theoretical performance of ACF using 3D synthetic seismic data additive white Gaussian noise. For comparison, we utilized three previously proposed seismic noise attenuation methods: damped rank reduction (DRR) \cite{25, 26}, hankel sparse low-rank approximation (HSLR) \cite{5}, and twice denoising autoencoder (TDAE) \cite{6}. 

\begin{figure*}
\centering
\noindent\includegraphics[width=12cm]{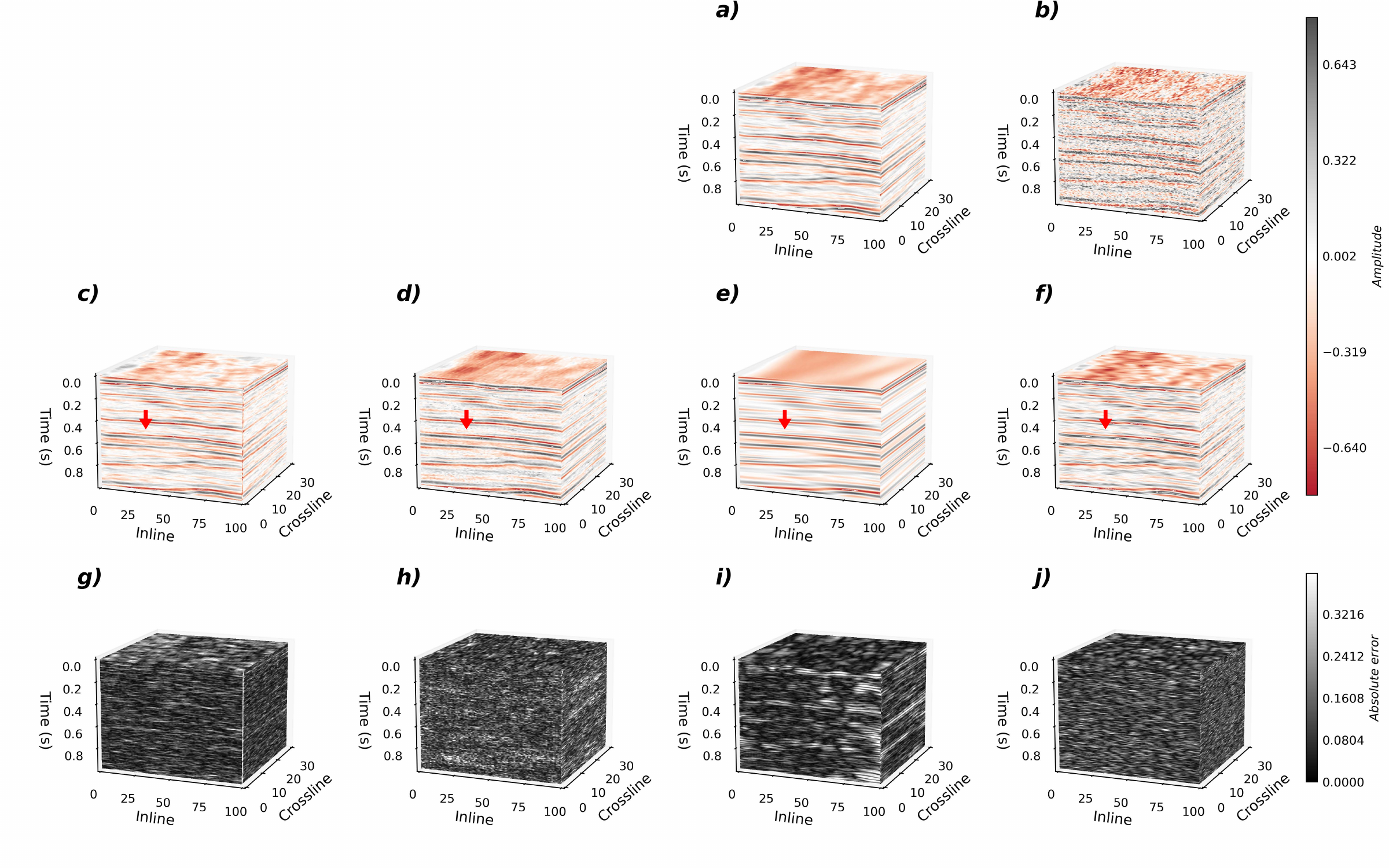}
\caption{3D synthetic seismic data. a) and b) represent the clean seismic data and the noisy seismic data, respectively; c) to f) represent the processed results of ACF, TDAE, DRR, and HSLR; g) to j) represent the absolute errors between clean data and results of ACF, TDAE, DRR, and HSLR.}
\label{synthetic data 2}
\end{figure*}
\begin{figure}
\centering
\noindent\includegraphics[width=9cm]{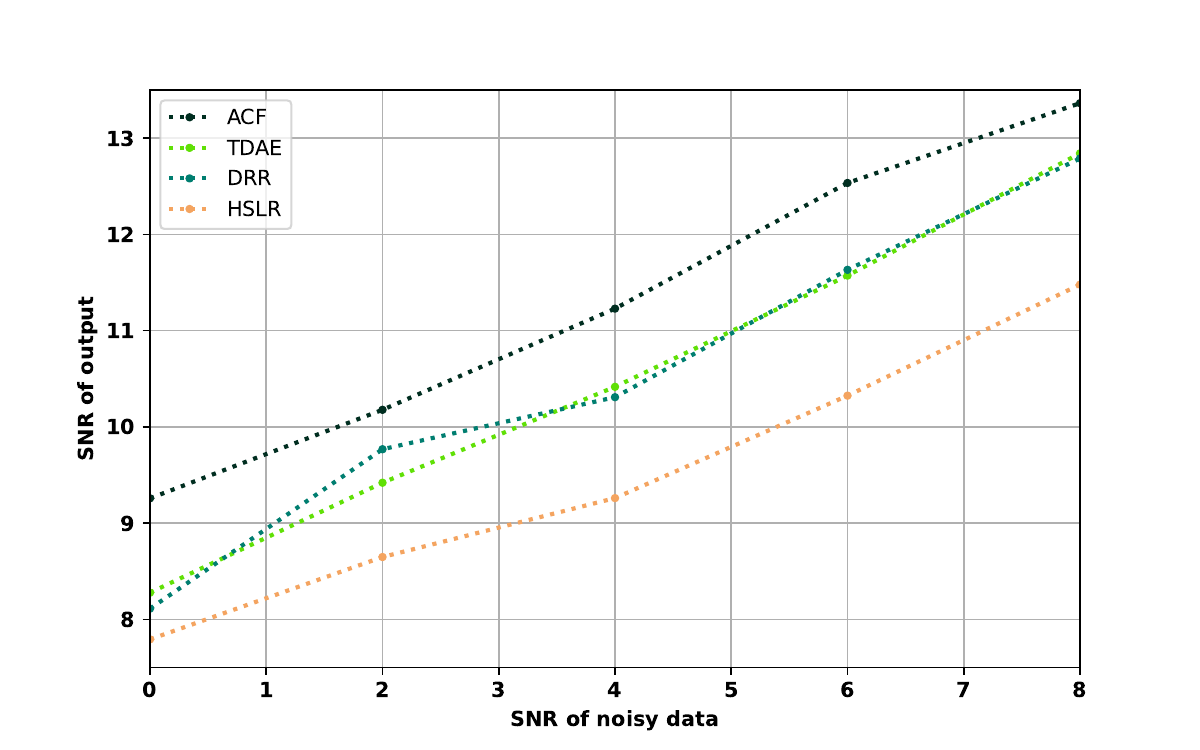}
\caption{SNR of four methods` results on synthetic data.}
\label{synthetic data metric}
\end{figure}

The synthetic data is shown in Figure~\ref{synthetic data 2}a. We conducted five sets of experiments with different noise levels, as shown in Table~\ref{synthetic data metric}. All methods can effectively attenuate noise, and ACF can achieve better results compared to other methods. Meanwhile, we show the results under strong noise conditions, as shown in Figure~\ref{synthetic data 2}b-f. In the processing results of the four methods, the result of ACF is closer to the ground truth (Figure~\ref{synthetic data 2}g) and its processed waveform is clear and can better restore the details of weak reflections (Figure~\ref{synthetic data 2}c). However, the drawback is that its boundary effect is obvious. The limitations of HSLR and TDAE are that HSLR's results contain substantial artifacts and TDAE's results retain residual noise, both of which reduce the SNR of their processing outcomes. The result of DRR is the smoothest. However, due to the high overlap between seismic data and noise, DRR struggles to attenuate noise while protecting effective signals, so there is a large deviation between its reflection signal and the ground truth (Figure~\ref{synthetic data 2}i). 

\subsection{Field Data Examples}

\begin{figure*}
\centering
\noindent\includegraphics[width=12cm]{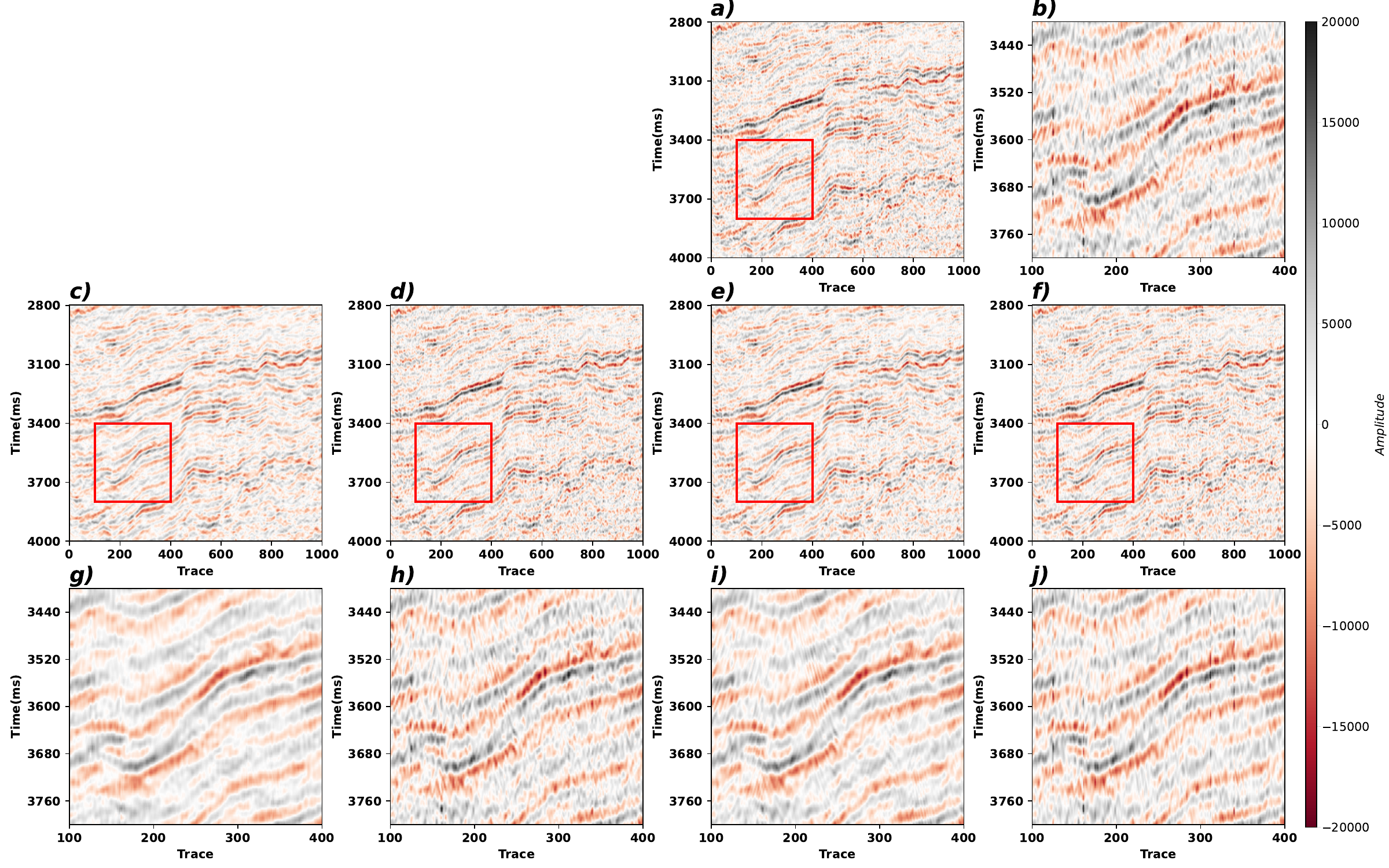}
\caption{Noisy field seismic data. a) represents the noisy field seismic data; b) represents a local enlarged view of the area within the red box; c) to f) represent the processed results of ACF, TDAE, DRR, and HSLR; g) to j) represent the local enlarged view of results of ACF, TDAE, DRR, and HSLR.}
\label{line1}
\end{figure*}

Since clean labels are typically unavailable for field data, we use average local similarity \cite{7} to assess the noise attenuation performance of various methods. The lower the local similarity, the better, which means that the correlation between the attenuated noise and seismic data is low. In addition, based on local similarity, we used the root-mean-square (RMS) noise intensity to evaluate the intensity of the removed noise, which can be represented as $RMS=\sqrt{\frac{\sum_{i=1}^n(I_i-O_i)^2}{n}}$. In this task, we adopted 2D seismic data for experiment, which was sourced from the National Petroleum Reserve–Alaska (NPRA) \cite{3}, as shown in Figure~\ref{line1}. It can be clearly seen that the processing results of ACF are smoother and the event horizons are more continuous compared to other methods.

\begin{figure}
\centering
\noindent\includegraphics[width=9cm]{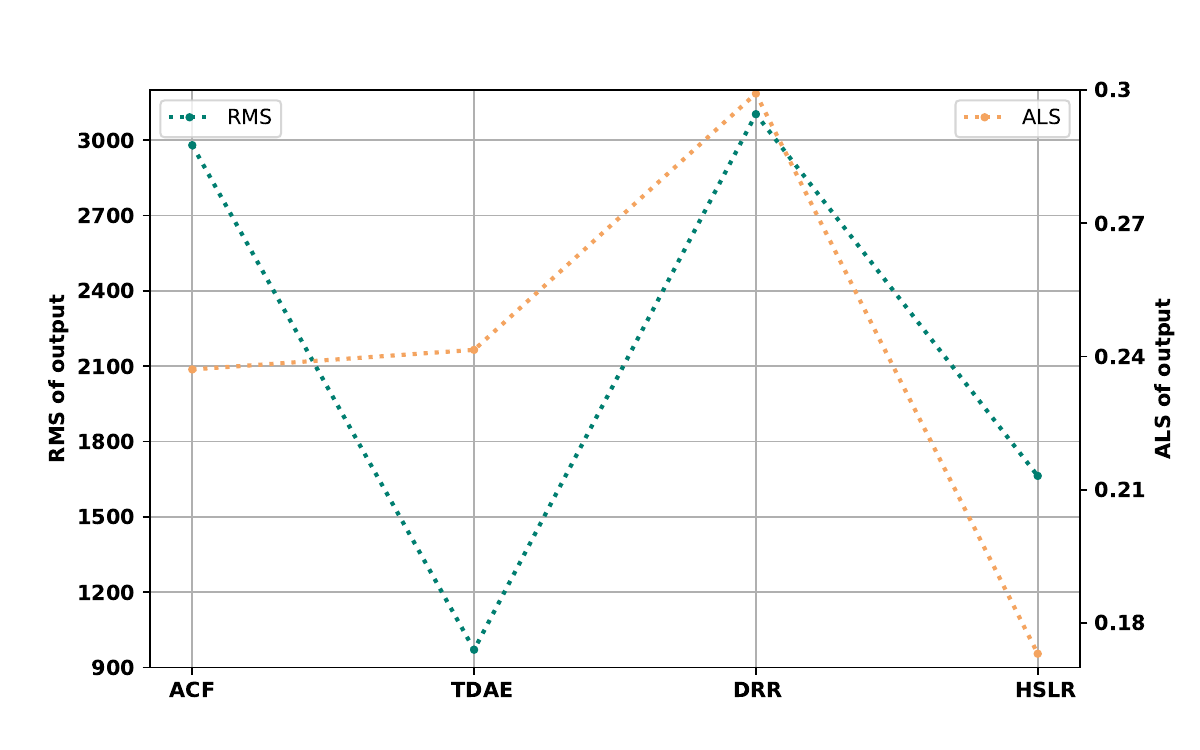}
\caption{RMS and ALS of four methods` results on field data.}
\label{field data metric}
\end{figure}

The results for the field data are shown in Figure~\ref{line1}. It can be clearly seen that the processing results of ACF are smoother and the event horizons are more continuous compared to other methods. Although the noise in field data is lower in frequency than Gaussian white noise, ACF still achieves good results, and we speculate that this is because low-scale learning effectively increases the frequency of the noise. However, other methods, without adopting such a strategy, still contain a large amount of noise or artifacts in their results (Figure~\ref{line1}h to j). It can be seen that TDAE and HSLR did not achieve satisfactory results, while DRR effectively attenuated most of the noise, but its results still contained a small amount of artifacts. To further compare the performance of various methods, we statistically analyzed the RMS and average local similarity (ALS) of each method (Figure~\ref{field data metric}). 

Due to the differences in the numerical ranges of seismic data, the RMS values of different datasets also vary. HSLR had the lowest ALS, indicating its excellent performance in handling high-frequency noise. However, its RMS was relatively low, suggesting that the noise attenuation performance of HSLR needs further improvement. TDAE also faces such a situation, struggling to effectively attenuate noise that is strongly aliased with signals. DRR and ACF showed similar performance, with DRR even capable of attenuating more noise (higher RMS). However, its ALS values were significantly larger, which can be attributed to signal leakage caused by excessive noise attenuation of DRR. In summary, ACF achieved a balance between ALS and RMS, while attenuating a significant amount of noise (high RMS), it ensured ALS second only to HSLR. 

\section{Conclusion}

We proposed a convolution filter ACF to improve the noise issue in seismic exploration caused by interference. ACF has only 2464 learnable parameters and relies on two priors: local prior and global variance prior. To further enhance the noise attenuation performance of ACF, we also propose a low-scale learning strategy. We used synthetic and field data for the experiments, and the results processed by ACF show that noise is effectively attenuated. Compared to three other comparative methods, ACF exhibits clearer waveforms of reflections from subsurface polygon, more continuous seismic event horizons, and higher resolution. Moreover, it maintains good computational efficiency and interpretability, which facilitates the direct application of ACF to practical seismic exploration tasks, effectively improving the accuracy of seismic exploration and helping us better understand the actual conditions of subsurface. 

\section{Acknowledgments}

This research is supported by the Chengdu University of Technology Postgraduate Innovative Cultivation Program. The codes of ACF have been uploaded to a GitHub repository \url{https://github.com/lexiaoheng/Mariana}.


\bibliography{cite}
\bibliographystyle{IEEEtran}
\end{document}